\documentclass[12pt]{iopart}

\begin{document}

\title{Finding passwords by random walks: How long does it take?}

\author{G. Kabatiansky$^{1,2}$ and G.Oshanin$^{2,3}$}
\address{$^1$Dobrushin Mathematical Laboratory,
Institute of Information Transmission Problems,
Russian Academy of Sciences,
Bolshoy Karetniy 19, Moscow GSP-4 101 447 Russia}
\address{$^2$Laboratory J.-V. Poncelet (UMI  CNRS 2615),
  Independent University of Moscow, Bolshoy Vlasyevskiy Pereulok 11, 119002 Moscow Russia}
\address{$^3$Laboratoire de Physique Th\'eorique de la Mati\`ere Condens\'ee (UMR CNRS 7600),
Universit\'e Pierre et Marie Curie, 4
place Jussieu, 75252 Paris Cedex 5 France}
\ead{kaba@iitp.ru; oshanin@lptmc.jussieu.fr}
\begin{abstract}
We compare an efficiency of a deterministic
"lawnmower" and random search strategies for finding a prescribed sequence of
letters (a password) of length $M$ in which all letters are taken from the same $Q$-ary alphabet. 
We show
that at best a random search takes two times longer than a "lawnmower" search. 
\end{abstract}
\pacs{05.40.Fb}
\vspace{2pc}
\noindent{\it Keywords}: random search, random walks, first passage times

\vspace{2pc}
\submitto{\JPA}

\section{Introduction}

Suppose one has forgotten a code or a password for
his multiple-dial combination lock (or any pin-protected electronic device).
Suppose next that the lock is perfect and is machined very precisely,
such that when any of the discs is being rotated, it does
not give any "click" or any other hint when a letter
or a numeral are at a correct position - this lock opens only when all the numerals or letters on all of the
discs form simultaneously a correct sequence.
How one should proceed in order to find a code?

An evident \textit{brute force} approach is to explore the space of
all possible combinations sequentially: starting from any
random combination, one rotates one of the discs completely, step by step,
from a symbol to a neighboring symbol,
then turns the second disc to a neighboring symbol,
rotates completely the first disc again, and etc. This procedure is repeated
until a correct sequence is found.

Let the desired code $\tilde{A}$ be
a sequence of $M$ symbols:
\begin{equation}
\tilde{A} = \{\tilde{a}_1, \tilde{a}_2, \tilde{a}_3, \ldots, \tilde{a}_M\},
\end{equation}
where each letter $\tilde{a}_m$ in the sequence
is taken from the same $Q$-ary alphabet $\{a\}$.
With such a "lawnmower" strategy,
given that a rotation of any of the discs
to the neighboring symbol takes one unit of time,
one is certain to find the desired
code within at most $N = Q^M$ time steps.
The probability $P_n$ that
the code is not
cracked up to  the $n$-th time step is given by
\begin{equation}
\label{1}
P_n = 1- \frac{n+1}{N}, \,\,\, n = 0,1, \ldots, N-1,
\end{equation}
while the probability $F_n$ that the code is first cracked exactly on the $n$-th step is $1/N$, such that within the "lawnmower" strategy the
mean first passage time $\overline{T_l}$ to the cracking event (or the expected life-time of the code) is simply
\begin{equation}
\label{l}
\overline{T_l} = \sum_{n = 0}^{N - 1} P_n = \frac{N - 1}{2} \sim \frac{N}{2}.
\end{equation}
The symbol $\sim$ here and henceforth signifies the exact behavior to leading order in $N$.

In this paper we pose a question how long it will take if, instead of a
sequential exploration of all possible combinations, we search for
the desired code in a random fashion. More specifically, our random search
algorithm is defined as follows: we first numerate the symbols in
 the alphabet $\{a\}$ and use numerals $0,1,2, \ldots , Q - 1$ instead of symbols. Then,
 at each tick of the clock we choose at random a numeral along the
 word and add to it either $+1$ or $-1$, independently on each step and with equal likelihood.
 At the next step, we choose again at random a numeral along the
 word and repeat the procedure. In original settings, it means
 that at each time step we choose at random a disc in our multiple-dial
 combination lock and rotate it downwards or upwards, with
 equal probability, to the neighboring symbol.
 Clearly, this process represents a nearest-neighbor
 random walk, commencing at a random site, on a periodic $M$-dimensional simple cubic lattice of linear size $Q$ and
 comprising $N = Q^M$ sites. The desired code $\tilde{A}$ can be thought of as some target site on this lattice.
 As in the case of a "lawnmower" search, we are interested to calculate 
 the probability that the code remains  not found until the $n$-th step, the
 distribution of the first-passage time
 to the target site and the expected life-time of the code.

 \section{Basic equations and results}

Let $a_m(n)$ denote the
value of the numeral at position $m$ along the word on the $n$-th time step and
$\delta(a)$ be the indicator function:
\begin{equation}
\label{ind}
\label{cases}
\delta(a)=\cases{1&for $a = 0$\\
0&for $a \neq 0$.\\}
\end{equation}
Then, the indicator function $I_n$ of the event that a given trajectory of a random walk
has not reached the target site $\tilde{A}$ within the first $n$ steps
can be written down as
\begin{equation}
\label{ind1}
I_n =  \prod_{n' = 0}^n \left(1 - \prod_{m=1}^M\delta(a_m(n') - \tilde{a}_m)\right),
\end{equation}
where
\begin{equation}
A(n') = \{a_1(n'), a_2(n'), a_3(n'), \ldots, a_M(n')\},
\end{equation}
denotes the random walker position on the lattice at time moment $n'$.

Averaging the expression in Eq.~(\ref{ind1}), we find that the probability that
the random walk has not reached the target site
up to time step $n$ is given by
\begin{equation}
\label{2}
P_n = 1 - \frac{S_n}{N},
\end{equation}
where $S_n$ is the expected number of
distinct sites visited by a random walk
on a periodic $M$-dimensional simple cubic lattice. We use here
the convention that $S_0 = 1$. Clearly, Eq.~(\ref{2}) is an analog of Eq.~(\ref{1}),
describing the form of $P_n$ within the "lawnmower" strategy.

Hence, the crucial property is $S_n$.
Explicitly, the expected number of distinct sites visited is determined as
\begin{equation}
S_n = \sum_{\tilde{A}} \left(1 - L_n(\tilde{A})\right),
\end{equation}
with $L_n(\tilde{A})$ being the probability that the simple random walk starting at the origin
at time moment $n = 0$ has not visited
the site $\tilde{A}$ up to the $n$-th step,
irrespective of the number of other sites it has visited till then.
Hence,
\begin{equation}
L_n(\tilde{A}) = 1 - \sum_{n'= 0}^n F_{n'}(\tilde{A})
\end{equation}
and
\begin{equation}
S_n = \sum_{n'= 0}^n \sum_{\tilde{A}} F_{n'}(\tilde{A}),
\end{equation}
where $F_n(\tilde{A})$ is the probability that the first visit
to the target site  $\tilde{A}$ occurred exactly on the $n$-th step \cite{katja,redner,1}.

Using the standard results on random walks properties
(see, e.g., Ref.\cite{1,2} and references therein), one finds
eventually the following general result:
\begin{equation}
\label{ss}
S_n = \frac{1}{2 \pi i} \oint \frac{d z}{z^{n + 1}} \frac{1}{(1 - z)^2} \frac{1}{G(0;z)}
\end{equation}
where the integral is around the origin of the $z$ plane and $G(0;z)$ is the generating function of the probability
to find the random walk at the origin at time $n$, given that it started at the origin at time $n = 0$,
\begin{equation}
\label{3}
G(0;z) = \frac{1}{N} \sum_{\bf q} \frac{1}{1 - z \, \lambda({\bf q})}.
\end{equation}
In Eq.~(\ref{3})
the function $\lambda({\bf q})$ is the structure function of the random walk:
\begin{equation}
\label{l}
\lambda({\bf q}) = \frac{1}{M} \left(\cos(q_1) + \cos(q_2) + \ldots + \cos(q_M)\right),
\end{equation}
while
${\bf q}$ is a $M$-dimensional vector with components $q_m = 2 \pi k_m/Q$, where $k_m = 0, 1, \ldots, Q - 1$ with $Q$ being the linear size of the lattice (length of the alphabet).

In what follows we focus on the situations when $M > 1$ and $Q \gg 1$. The case $M = 1$ corresponds to Brownian search in one-dimensional systems and has been extensively discussed recently in view of possible improvements by, e.g., intermittent random walks \cite{olivier,we,we1}. The case of binary alphabets with $Q = 2$ describes an interesting case of search in the Hamming space and will be discussed elsewhere \cite{we2}.

Consider now the the form of $P_n$ in Eq.~(\ref{2}). For sufficiently small $n$
each new visited site is most likely a "virgin" site \cite{1}, i.e., a site visited for the first time. Hence, at short times $S_n \sim n$ and $P_n$ in Eq.~(\ref{2}) exhibits essentially the same behavior as its counterpart in Eq.~(\ref{1}), describing the efficiency of the "lawnmower" search.
Similarly, at short times the probability $F_n$ that the code is cracked for the first time exactly on the $n$-th step is $1/N$.

At greater times, however, the growth of $S_n$ saturates and $S_n$ approaches $N$ - the total number of different combinations. The relaxation
of $S_n$ to its ultimate value $S_{\infty} = N$ is an exponential
function of the form
\begin{equation}
\label{s}
S_n \sim N \left(1 - \exp\left(- \frac{n}{\tau}\right)\right), \,\,\, {\rm as} \,\, n \to \infty,
\end{equation}
where $\tau$ is
the largest relaxation time.
Calculation of
 $\tau$ is
a rather delicate mathematical problem and we address the reader
to Ref.\cite{2} for more details. It was shown in Ref.\cite{2} that for sufficiently large $Q$,
\begin{equation}
\label{tau}
\label{cases}
\tau = N \, \cases{G &for $M \geq 3$\\
\ln(c N)/\pi &for $ M = 2 $,\\}
\end{equation}
where $G$ and $c$ are constants: $c \approx 1.8456$, while $G$
is given by an $M$-fold integral
\begin{equation}
G = \frac{1}{\pi^M} \int_0^{\pi} \ldots \int_0^{\pi} \frac{\prod_{m = 1}^M d x_m}{1 - \lambda({\bf x})}
\end{equation}
with $\lambda({\bf x})$ defined by Eq.~(\ref{l}) (with the replacement $q_m \to x_m$). One notices that $G$ is just the mean number of visits
to the origin by standard nearest-neighbor random walk, commencing at the origin, on a $M$-dimensional infinite simple cubic
lattice within an infinite time.

Therefore, in the large-$n$ limit, we get, in virtue of Eqs.~(\ref{2}) and (\ref{s}), that
\begin{equation}
P_n \sim \exp\left(- \frac{n}{\tau}\right),
\end{equation}
and hence, since $F_n = P_n - P_{n+1}$, the first passage time distribution 
$F_n$ has also an exponential tail with the characteristic decay time $\tau$.

The mean first passage time $\overline{T_r}$ to the cracking event or the life-time of the code
can be determined exactly from Eqs.(\ref{2}) and (\ref{ss}), $\overline{T_r} = \sum_{n = 0}^{\infty} P_n$. It appears that
$\overline{T_r}$ \cite{mont2} coincides with the largest relaxation time $\tau$, Eq.(\ref{tau}). Comparison of $\tau$, Eq.(\ref{tau}), and of $\overline{T_l}$ in Eq.~(\ref{l}) allows us to draw the following conclusions:
\begin{itemize}
\item For this problem the "lawnmower" search always outperforms a "random" search algorithm.
\item The worst performance of a "random" search is for "two-letter" codes since here the mean first passage time $\tau$ contains an additional logarithmic factor $\ln(N)$ compared to the "lawnmower" result.
\item For three (and longer) letter codes the mean first passage time $\tau$ scales linearly with $N$, i.e. exactly as $\overline{T_l}$ does. However, $\tau$ is always larger than  $\overline{T_l}$ due to a numerical factor $f = 2 G$. $G$ is a decreasing function of the code length; for example, for three-letter codes $G \approx 1.516$, for four-letter codes $G \approx 1.239$, for five-letter codes $G \approx 1.156$ and etc. For larger $M$, the following asymptotic expansion holds \cite{mont1}:
\begin{equation}
\label{exp}
G = 1 + \frac{1}{2 M} + \frac{3}{4 M^2} + \mathcal{O}\left(\frac{1}{M^3}\right).
\end{equation}
Hence, the ratio $\tau/\overline{T_l} \to 2$ when the length of the code increases; it thus takes at best two times longer
to crack a code using a random search than within the "lawnmower" search.
\end{itemize}

Finally, we discuss a little bit different random algorithm in which, after choosing at random a numeral in the code, we increment it with equal likelihood by $\delta = \pm 1, \pm 2, \pm 3, \ldots ,\pm l$. It means that after having chosen a disc, we turn it upwards or downwards on any integer
distance within an interval $[1,l]$.
Clearly, for such an algorithm all the results in Eqs.~(\ref{2}) to (\ref{3}), as well as Eqs.~(\ref{s}) and (\ref{tau}), still hold, except for the definition of $\lambda({\bf q})$. In this, more general case, the structure function of the random walk is given by:
\begin{equation}
\label{ll}
\lambda({\bf q}) = \frac{1}{l \, M} \sum_{m = 1}^M \sum_{j = 1}^l \cos(j \, q_m),
\end{equation}
while $\tau$ is defined by Eq.~(\ref{tau}) with
\begin{equation}
G = G_l = \frac{1}{\pi^M} \int_0^{\pi} \ldots \int_0^{\pi} \prod_{m = 1}^M d x_m \, \left(1 - \frac{1}{l \, M} \sum_{m = 1}^M \sum_{j = 1}^l \cos(j \, x_m)\right)^{-1}.
\end{equation}
Some straightforward analysis shows that $G_l$ is a monotonically decreasing function of $l$. One readily finds an expansion similar to the one in Eq.~(\ref{exp}),
\begin{equation}
G_l \approx 1 + \frac{1}{2 M l}.
\end{equation}
Hence, such a random algorithm appears to be more efficient, for large $l$,
than the $l = 1$ case and $G$ can be made very close to unity for any $M$.
On the other hand, this algorithm can not outperform the "lawnmower" search and within the former it will
take at least two times longer to find a code compared to the latter one.

\section{Conclusions}

To conclude, we have compared an efficiency of a deterministic
"lawnmower" and of random search strategies for finding a prescribed sequence of
letters - a password - in words of length $M$ with letters taken from the same $Q$-ary alphabet. We have shown
that at best a search within a random strategy takes two times longer than within a "lawnmower" search.

We note that the search
of a password - a given sequence of letters -
in the sequence space can be viewed as a (random) walk on a single-connected graph. Here, each node of the graph corresponds to a particular configuration of the lock while each  bond corresponds to a physically possible one-step
transformation of the lock. Clearly that for any such graph
possessing a Hamiltonian cycle, the "lawnmower" search for a random target site outperforms
random search. The question is in how many times? Graphs considered in this paper are examples of strongly regular graphs \cite{bose}, and we suppose that in a general case the answer for the question can be done in terms, for instance, of the eigenvalues of the graph.

We finally remark that the
 problem discussed here
 can be viewed from a different perspective (see \cite{gros} for more details).
 Suppose one has a polymer containing $M$ monomeric units,
 and each of these units can be of $Q$ different types.
 Starting from a particular sequence, one allows then for mutations of
 the monomers from one type to another. The "goal" of the polymer
 is to attain some specific ("foldable" in \cite{gros}) configuration.
In terms of our model, this process represents a random
search algorithm in which rotation
of any of the
discs on an arbitrary distance is allowed and several discs can be rotated simultaneously.

\section{Acknowledgments}

 We acknowledge helpful discussions with A.Yu.Grosberg and also wish to thank him for pointing us on the analogies presented in Ref.\cite{gros}.
 G.O. is partially supported by Agence Nationale de la Recherche
(ANR) under grant ``DYOPTRI - Dynamique et Optimisation des Processus de
Transport Intermittents''.

\end{document}